\documentclass[]{article}

\usepackage[fontsize=9]{scrextend}
\usepackage{mathptmx}

\usepackage[a4paper, total={6.75in, 10in}]{geometry}
\usepackage[table]{xcolor}
\usepackage{fancyhdr}
\usepackage{hyperref}

\usepackage{graphicx}
\usepackage{multicol}
\usepackage{array,booktabs}
\newcolumntype{P}[1]{>{\centering\arraybackslash}p{#1}}

\usepackage{float}
\usepackage{amsmath}

\usepackage[style=apa,citestyle=authoryear,backend=biber]{biblatex}
\usepackage{sectsty}
\sectionfont{\normalfont\fontsize{9}{9}\selectfont}
\subsectionfont{\normalfont\fontsize{9}{9}\selectfont}
\subsubsectionfont{\normalfont\fontsize{9}{9}\selectfont}

\usepackage{blindtext}

\hypersetup{
    pdfborderstyle={/S/U/W 1}, 
    colorlinks,
    linkcolor={red!50!black},
    citecolor={blue!50!black},
    urlcolor={blue!80!black},
    pdftitle={A 3D Mesh Based Approach to In Home Safe Walking Spaces},
    pdfsubject={International Conference on Smart Education and ICT},
    pdfauthor={Aaron S. Crandall, PhD},
    pdfkeywords={Gerontechnology, falling, senior care, tripping hazards, augmented reality}
}

\usepackage{tikz}

\begin{document}

\begin{center}
    \begin{tabular}{ p{0.97\linewidth} }
        \multicolumn{1}{c}{\Large A 3D Mesh Based Approach to In Home Safe Walking Spaces} \\
        \multicolumn{1}{c}{\Large for Older Adults} \\
        \\
        \\
        {\large \textbf{Aaron S. Crandall}} \\
        {\large Gonzaga University, United States of America} \\
        crandall@gonzaga.edu \\
    \end{tabular}
\end{center}

\vspace{0.3in}

\noindent
ABSTRACT: Falling continues to be a significant risk factor for older adults and other mobility limited individuals.
Monitoring and maintaining clear, tripping hazard free pathways in living spaces is invaluable in helping people live independently and safely in their home.
This paper proposes and demonstrates a Microsoft HoloLens-based technology for monitoring the open walking areas in living spaces.
The system is based on the 3D mesh produced by the HoloLens augmented reality device.
Several algorithms were implemented and evaluated to handle the noisy raw mesh to help identify all possible floor spaces.
The resulting 3D model of the home was then processed to find United States Occupational Safety and Health Administration (OSHA) approved pathways through the homes. 
The long term goals of these technologies are to monitor the living space’s clutter and clear walkways over time.
The information it generates shall be provided to the residents and their caregivers during environmental home assessments.
It will inform them about how well the home is being maintained so proactive interventions may be taken before a fall occurs.

\vspace{0.1in}

\noindent
R\'{E}SUM\'{E}: Les chutes continuent d'\^{e}tre un facteur de risque important pour les personnes \^{a}g\'{e}es et les personnes \`{a} mobilit\'{e} r\'{e}duite.
La surveillance et le maintien de passages d\'{e}gag\'{e}s ne pr\'{e}sentant aucun risque de tr\'{e}buchement dans les espaces de vie sont indispensables.
Ces derniers aident les personnes \`{a} vivre de mani\`{e}re autonome et en toute s\'{e}curit\'{e} dans leur maison.
Cet article propose et pr\'{e}sente une technologie bas\'{e}e sur Microsoft HoloLens permettant de surveiller les zones de passage dans les espaces de vie.
Le syst\`{e}me fonctionne sur la base du maillage 3D produit par le dispositif de r\'{e}alit\'{e} augment\'{e}e HoloLens.
Plusieurs algorithmes ont \'{e}t\'{e} d\'{e}velopp\'{e}s et \'{e}valu\'{e}s pour traiter le maillage brut parasit\'{e} et aider \`{a} identifier tous les espaces au sol \'{e}ventuels.
Le mod\`{e}le 3D de l'habitation ainsi cr\'{e}\'{e} est ensuite trait\'{e} pour d\'{e}terminer les passages approuv\'{e}s par l'Occupational Safety and Health Administration (OSHA) des \'{E}tats-Unis.
Les objectifs \`{a} long terme de ces technologies sont de surveiller l'encombrement de l'espace de vie et de d\'{e}gager les passages au fil du temps.
Les informations qu'il g\'{e}n\`{e}re sont fournies aux r\'{e}sidents et \`{a} leurs auxiliaires de vie lors des \'{e}valuations environnementales du domicile.
Il les informera de la qualit\'{e} de l'entretien de la maison, ce qui permettra d'intervenir de mani\`{e}re proactive avant qu'une chute ne se produise.

\vspace{0.3in}


\vspace{0.1in}

\begin{multicols}{2}

  \section{INTRODUCTION}

According to the Centers for Disease Control, around 15.9\% of all US adults age 65 and over fell at least once in a 3 month window~\parencite{Stevens2008}.
These falls affect one in three adults over the age of 65 annually and 50\% of adults over the age of 80~\parencite{Ambrose2013}.
The cost of these falls accounts for 0.1\% of all healthcare costs in the United States annually and up to 1.5\% of healthcare costs in European countries.
The number one reason older adults have increased care needs is due to falls and fall related injuries.

The home is the most common location for seniors experiencing falls.
Around 55\% of all fall related injuries occur in the home~\parencite{Pynoos2010}, and 35\% to 40\% of those falls are directly related to environmental factors.
New approaches to preventing, detecting, and assisting in fall management are of utmost importance to the clinical community.

During prior work by the authors in gathering clinical needs for senior care focused smart home technologies, nurses often cited the need for help with several common topics.
These included medicine compliance, falls, and diet~\parencite{Zulas2014,Zulas2014a}.
When it came to falls, nurses normally focused on two major targets: clutter on the floors and hoarding.

\begin{quote}
  ``Fall risk.
  There are many things that nursing staff indicated were risk factors for falls that might be able to be detected more easily than falls themselves.
  Nursing staff also indicated that often it was \textbf{better to prevent falls than to deal with a fall once it occurred}.
  Risk factors for falls included clutter in the homes, ... .
  Likely, \textbf{detecting these things early would better help prevent falls}, allowing individuals to stay independent and safe for longer.'' (emphasis added)
\end{quote}

Additionally, family members of seniors spoke of dealing with tripping hazards, in-home walking safety, and needing to keep the home orderly~\parencite{Zulas2012}.
Having more information about how seniors are living is a key component of providing quality long term aging in place care~\parencite{Tyrer2006, Kaye2008}.
This leads to an opening for smart home and sensor technologies to gather data, analyze, model, and report to caregivers about a senior and their status.
The overall goal of in-home gerontechnology tools is to provide actionable information to caregivers to improve both independence of the seniors and the quality of care given.

While the field of fall prevention is well studied, most of the medical focus has been on clinical, medicine, and assistive tools.
These are well summarized in by the American and British Geriatrics Society's works on the topic~\parencite{Society2001}.
The area that receives a relatively small amount of work is in tools for facilitated environmental home assessments.
A home assessment is performed by a caregiver, clinician, or care facility manager to identify tripping hazards such as loose rugs, lack of grab bars in high risk locations, and overly narrow walking paths throughout the home.
In an ideal world, an assessment would be performed regularly, though that is often cost prohibitive.

This work proposes and implements a part of the home assessment process through automatic means.
To address the problem of clutter and safe walking spaces, the authors implemented a system centered around in-home 3D models gathered through sensor systems.
In this case, a HoloLens was used to gather a 3D mesh of the home.
This 3D mesh provides a spatial mapping of a home through a combination of infrared, vision, and depth mapping algorithms.
Under good conditions, the HoloLens mesh is accurate to centimeters allowing the detection of objects that could be classified as tripping hazards around the room~\parencite{Khoshelham2019}.
This mesh can then be interpreted for a combination of safe walking spaces and likely hazards.

Ideally, this kind of sensor system could be deployed and run often in a home, such as part of a larger smart home system~\parencite{Crandall2012}.
The net result would be a history of how clean and orderly the home's walking spaces are.
This information could be leveraged by caregivers to determine if interventions are needed to help remove tripping hazards from an older adult's home.
Over time, this approach should enable a more mobile and safe environment for people to age in place.

This work summarizes the implementation of a system centered around identifying safe walking spaces in homes.
It uses several algorithms to handle the complexity and noise in the HoloLens 3D mesh output.
The results are analyzed for safe paths based around the US Occupational Safety and Health Administration suggested safe walking path regulations~\parencite{OSHA2021}.
In this case, it means that a path needs to be at least 0.46m (18 inches) wide for safe passage.
This value can easily be changed to analyze the room for the 0.91m (36 inches) required by the Americans with Disabilities Act~\parencite{ADA2002} if the older adult is more mobility impaired or needs more room for assistive tools such as a walker or wheelchair.
Additional analyses can be added for other safety standards regarding walking spaces, doorways, and stairs.

  \section{Related Works}
\label{sec:related_works}

Addressing the issues of falls and falling has been an ongoing effort in the research community.
The range of approaches includes critical situation detection, early detection, and preventative.
All of these works have a place during the arc of care.

Examples of critical situation detection include fall detection such as Skubic's early work on detecting falls via video~\parencite{Skubic2009}.
More modern works leverage deep learning~\parencite{Islam2020}, neural networks~\parencite{Yao2020}, and incorporate privacy protecting algorithms for real world deployments~\parencite{Liu2021}.

These fall detection systems use a variety of sensors to detect critical events.
The sensors could include wireless radio fields~\parencite{Ding2020}, microphones~\parencite{Li2012}, accelerometers~\parencite{AlNahian2021}, or floor vibration sensors~\parencite{Shao2021}.
All of these approaches demonstrate various degrees of success, though with trade offs in their deployability, cost, detection rates, and false negatives.

A more proactive-oriented field of work is in fall prediction.
In these cases, the research focuses on predicting the likelihood of a fall by an older adult.
The systems usually use a similar suite of sensors, though the algorithms are centered around prediction instead of detection of an event.
The use of accelerometers in smart phones and watches is common given their deployability and other uses to users~\parencite{Tai2020}.
One advantage of these tools is they can use posture~\parencite{Howcroft2017} and gait~\parencite{Buisseret2020} to help with fall risk classification.
These kinds of approaches are valuable to clinicians as a source of additional information about patients and wards in their care.
The models are also often built around existing methods of observation such as the Timed Up \& Go test~\parencite{Podsiadlo1991} or Sit-To-Stand transitions, which are well established clinical tools.

There is a relative dearth of work on automatic environmental home assessment, though there are semi-autonomous smart home approaches such as clinician-in-the-loop~\parencite{Fritz2017}.
The technologies required to perform 3D model based assessments are very new.
This work addresses part of this needed area of information gathering, analysis, and presentation surrounding home safety.
The project reported here is an extension on earlier works including:~\parencite{Phillips2017,Crandall2018,Crandall2020}.

  \section{Methods}
\label{sec:methods}


To build and evaluate a system for finding safe walking spaces, several system components were designed and used.
These components included a sensor package on a HoloLens version 1, an augmented reality interface, and a Linux server with a database and web server.

The sensor source was a Microsoft HoloLens version 1.
This device runs a Unity AR application to start, stop, and upload mesh data from the home.
The application uses SFTP to copy a home's data to the Linux server for storage.



The server can be broken down into three subsystems: data storage, Walking Spaces Algorithm (WSA), web services.
The data storage subsystem manages all of the data movement and user notifications.
The WSA tools do the primary parsing, analysis, and generation of information for each 3D mesh uploaded by the HoloLens.
The website offers a way for the end-user to view and interact with the data.

The web system provides a back-end API for communication with the MongoDB storage archive.
The front end is written in React and provides a rendering of the 3D meshes for users to review and interact with.


Data storage is managed by several Python programs running on the Linux server.
These programs watch for newly uploaded 3D mesh object files copied via SFTP to the server from the HoloLens.
They then put a copy of the object file into a long term archive for future research work as well as inserting them into the MongoDB server.
Once archived, the data storage program starts the WSA analysis of the new mesh.


\begin{figure}[H]
  \centering
  \includegraphics[width=0.75\columnwidth,keepaspectratio]{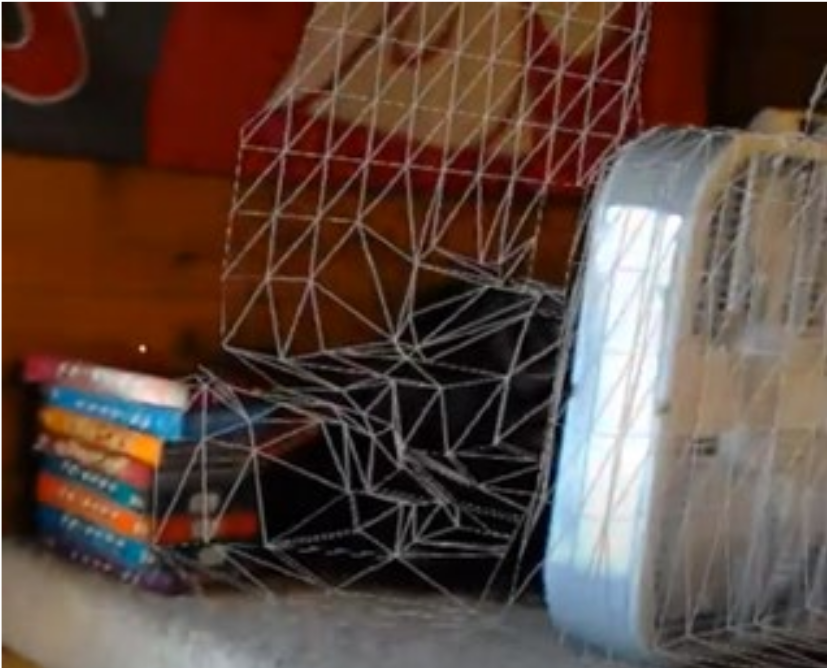}
  \caption{3D mesh overlayed on objects in the room.}
  \label{fig:mesh_over_objects}
\end{figure}

For this project, all 3D meshes are stored in the Wavefront object file format (.OBJ), which is readily usable by Unity on the HoloLens~\parencite{Lee2019}.
This data provides vertex data from the estimated surfaces discovered by the HoloLens as it scanned the home.
The values include geometric vertices, texture vertices, vertex normals, and parameter space vertices.
Together, these points provide a set of triangular surfaces that roughly correspond to the 3D space.
Only the vertices and resulting triangular surfaces are kept for this project to limit file size and provide compatibility with existing libraries used in analysis.
The OBJ format also allows to group vertices into named sets, which becomes valuable when annotating and classifying the 3D mesh after running the WSA tools.
An example of the 3D mesh overlayed on some objects scanned by the HoloLens is shown in Figure~\ref{fig:mesh_over_objects}.


The WSA algorithms generate meshes for different subsets of the rooms in the 3D mesh.
Each of these meshes are designed to identify particular pieces of interest to a user.
Notably, outputs include safe walking spaces, raised working spaces like tables, and identified clutter on the floors.

Based on these mesh outputs we are able to identify a subset of the total estimated floor space that pass OSHA requirements as a safe walkway.
These regions can then be highlighted with various colors to show degrees of walkable spaces.
The highlighted images are stored on the web server for the user to access and review at a later time.

Processing the initial OBJ file has several steps to reduce noise and unnecessary information.
A first step is to remove the ceiling and all mesh above 2m in the room.
These data points will not be part of a walking space or clutter on the floor.

To identify the likely floor, all triangles are analyzed for their normal to the Z axis.
Triangles with a normal with a divergence greater than 1 degrees ($min_\Theta$) are removed.
Of the remaining set, the triangles are clustered into groups representing likely bands of true floor height.
The triangles in the lowest height group are averaged and any triangles below that average are considered noisy and removed.
Of the remaining triangles, they represent a highly accurate elevation for the primary floor height within the room.

Given the likely floor height, a ground plane is established and the mesh can be subdivided into open floor spaces and non-open spaces.
This is done by a waterfall algorithm~\parencite{Hanbury2006}.
''Droplets'' of simulated water are dropped from the highest elevation in the 3D space occupied by the mesh.
Any droplets that reach the ground level without touching a triangle on route are considered to have reached a clear floor space.
The triangles in that droplet's landing zone are considered a clear walking space and included in the clear walking space mesh.
Any droplets that stop before reaching the ground plane add the triangle they impacted into the ''other objects'' category.

This process is repeated for higher elevation flat surfaces to detect tables and elevated working spaces.
These spaces are not walking areas, but are analyzed for clutter independently of the floor.

The edges of the walking spaces are also classified as either clutter or larger objects by comparing all triangles who are members of the floor group with their neighbors.
The class of a given neighbor is determined by first calculating the $\delta_\Theta$ between floor and neighbor as shown in Equation~\ref{eqn:dt}.
The neighboring tile is then classified according to Equation~\ref{eqn:classify}.
These additional classifications are kept in the OBJ file for rendering in the resulting images for the user.

\begin{equation}
    \delta_\Theta = abs(floor\_angle - neighbor\_angle)
    \label{eqn:dt}
\end{equation}

\begin{equation}
    neighbor\_class = \begin{cases}
        clutter, &  \delta_\Theta > min_\Theta \text{ and } \delta_\Theta < max_\Theta\\
        furniture, & otherwise.
    \end{cases} \\
    \label{eqn:classify}
\end{equation}

\vspace{3mm}

Post WSA algorithm completion, the server has multiple OBJ files, each with subsets of the original 3D mesh.
These subsets are viewable individually to see the calculated walking spaces, raised work surfaces, calculated clutter in the room, and other objects.
The resulting walking spaces mesh is then analyzed for valid OSHA walking space open areas.
For this work, each 0.15$m^2$ is processed as the center of a circle as a potential open walking space.
If that point's circle is entirely on the flat floor area with a radius of 0.46m (18 in), then it is colored green in the final visual renderings.
Any location that is on the floor, but does not have a full 0.46m circle around it of open space is colored yellow, and if it only has 10cm or less it is colored red.
This quickly highlights what is an OSHA approved pathway or not to a viewer.


The Walking Spaces system was evaluated in two homes, a lab at a university, and a student apartment.
In every case the 3D mesh was collected and archived for later processing.
The results were reasonable in each case, though the HoloLens has significant trouble with darker surfaces, windows, mirrors, and small objects under about 3" in height.
Future systems, both through sensor and 3D mesh creation improvements should reduce these limitations.


This approach has several limitations and assumptions.
The current algorithms process a home as a single unit, so individual rooms are not segmented for notation or exclusion by the user if they're not to be included in an analysis.
Stairs are not considered since all homes and spaces used were single story locations.
The Unity application used to do the collection was designed for simple sampling and not continuous collection, so the scale of data or processing requirements for a home were not evaluated.


Evaluation of the algorithms used was done by inspection by a researcher at all stages of processing.
The researcher used photos taken from the homes as ground truth.
The researcher was able to inspect the resulting mesh's match to floor, clutter, raised surfaces, and walking spaces for accuracy.

The data collection process also measured open floor spaces where a simple map was drawn of the space and main walking space widths were marked.
These marked up maps allowed the researcher to identify where the algorithms were reasonably correct or not.

  \section{Results}
\label{sec:results}

The WSA algorithm and resulting visualizations presented to the user on the web site show important details about the living spaces scanned and processed.
A caregiver, nurse, family member, or the residents themselves can readily see if there are safe walking paths through the home.
The resulting 3D meshes should also be analyzable by a simple shortest path algorithm such as A* to determine if the resident is able to safely move from key locations such as the bedroom to the bathroom through OSHA or ADA compliant routes.

\subsection{WSA Algorithm Analysis and Results}

\begin{figure}[H]
    \centering
    \includegraphics[width=0.75\columnwidth,keepaspectratio]{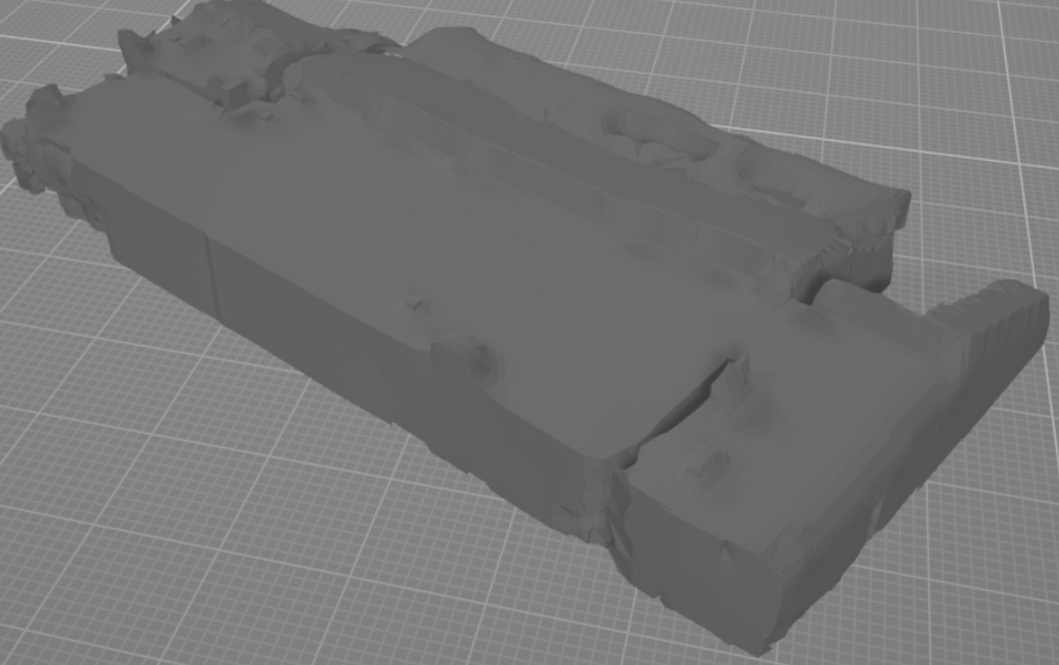}
    \caption{3D mesh for Home \#1 including ceiling data.}
    \label{fig:home1_full_data}
\end{figure}

An example result for one of the homes scanned is shown here.
The first image is the 3D mesh for the whole home, including the ceiling information.
The image in Figure~\ref{fig:home1_full_data} shows how the HoloLens fills in ceiling data, which can be mistaken for a floor given how flat it is.

The next images show the various subsets of the mesh as output by the WSA algorithm.
These results are taken from Home \#1, as recorded by the researchers.
The results from the other home, lab, and apartment were very similar.

\subsubsection{WSA Floor Mesh Output}

\begin{figure}[H]
    \centering
    \includegraphics[width=0.75\columnwidth,keepaspectratio]{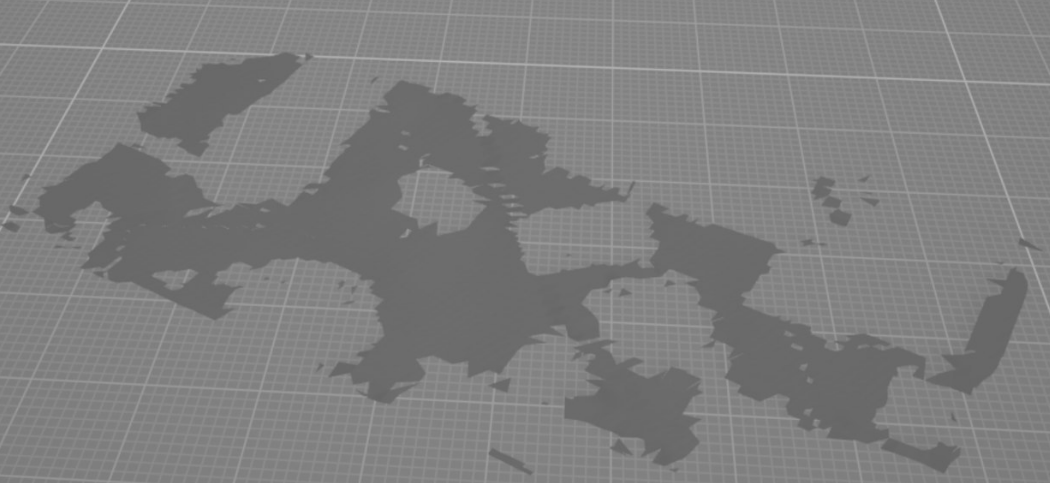}
    \caption{WSA floor mesh output for Home \#1.}
    \label{fig:home1_floor_mesh}
\end{figure}

The mesh shown in Figure~\ref{fig:home1_floor_mesh} is the floor of Home \#1.
These are the OBJ file triangles found to be oriented normally to the ground plane with noisy triangles removed.
The large open areas missing from the mesh are furniture, cabinets, walls, clutter, and a few areas not detected well by the HoloLens.

One area the algorithm did not perform well in for the floor discovery is in doorway thresholds.
These areas were often not recorded accurately by the HoloLens, and the resulting noise would not be handled by the WSA filters well.
The gap in the floor mesh on the left hand side is an extreme example of this error where a whole room is disconnected from the rest of the house.
Additional work is needed to improve on this behavior to more accurately represent the available walking spaces.

\subsubsection{WSA Clutter Mesh Output}

\begin{figure}[H]
    \centering
    \includegraphics[width=0.75\columnwidth,keepaspectratio]{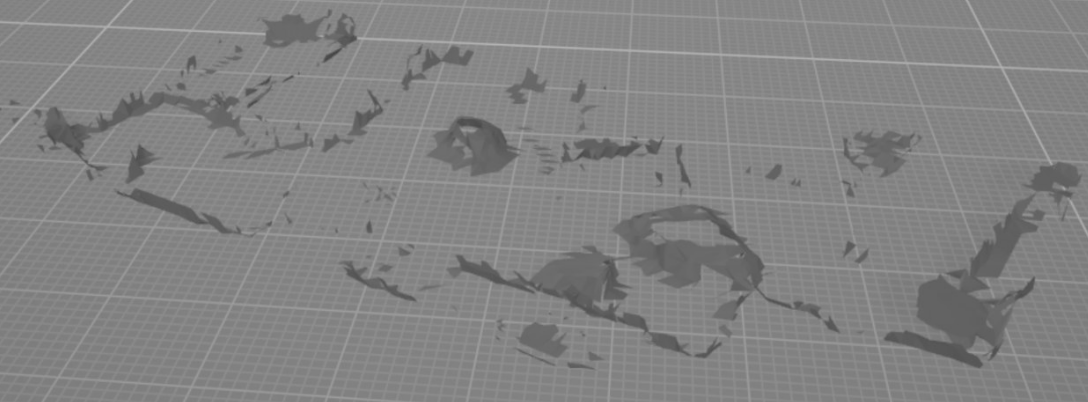}
    \caption{WSA floor clutter output for Home \#1.}
    \label{fig:home1_floor_clutter}
\end{figure}

The clutter objects discovered by the WSA are shown in Figure~\ref{fig:home1_floor_clutter}.
In this home, there were several piles of objects around the edges of many rooms, and especially the entry hallway on the right.
These piles were made of books, clothes, and hobby items.
The WSA did identify many of the these piles properly, though there were several regions of false positives.

False positives for clutter mainly occurred where the 3D mesh emitted by the HoloLens did not track the floor to wall transition very tightly.
The triangles in the mesh at this location were sitting at an angle just below our threshold for $max_\Theta$ in Equation~\ref{eqn:classify}.
This problem was accentuated in low light or if the wall was a darker color, which is a known limitation of the HoloLens v1 sensor platform.

The researcher was able to identify many of the objects classified as clutter directly from the photos taken in the home.
The algorithm agreed with the human classification most of the time, though more work is needed to generate more quantified results in future work on the degree of accuracy.

\subsubsection{WSA Work Surface Mesh Output}

\begin{figure}[H]
    \centering
    \includegraphics[width=0.75\columnwidth,keepaspectratio]{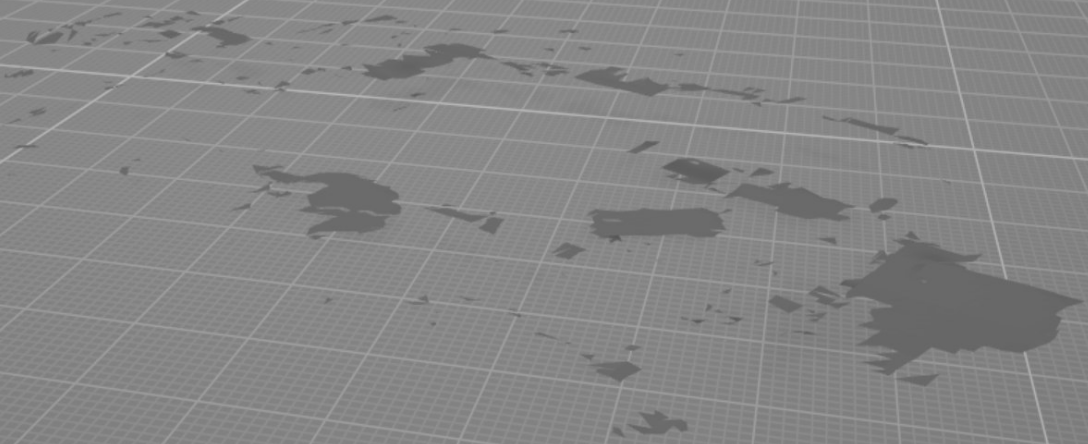}
    \caption{WSA work surfaces output for Home \#1.}
    \label{fig:home1_work_surface}
\end{figure}

The WSA tools emit a mesh of the raised work surfaces found in the home, as shown in Figure~\ref{fig:home1_work_surface}.
These are primarily tables, desks, kitchen counters, bathroom counters, nightstands, and other raise flat surfaces.
The researcher was the able to quickly identify the home's kitchen, a desk, the dining room table, and a coffee table from the mesh output for home \#1.
The other home and apartment had similar behaviors to the ones shown here and the lab space was even clearer with larger raised surfaces.

The large area kept in the raised work surface mesh on the right is a combination of a bed and a table in the home's bedroom.
The HoloLens made the bed a relatively flat, uniform surface so the WSA grouped it into a single object with the work surfaces.

These raised surfaces are analyzed for clutter, just as the floor is.
The clutter on the raised surfaces is marked in the final renderings to show how cluttered the work surfaces in the home are.
The raised surface itself is marked green to show how usable the workspace is.
Future work will address how to factor in the overall clutter of the home as it changes over time, both the floor and raised surfaces as a measure of independent living and home upkeep for older adults.

\subsubsection{WSA Home 1 Other Mesh}

\begin{figure}[H]
    \centering
    \includegraphics[width=0.75\columnwidth,keepaspectratio]{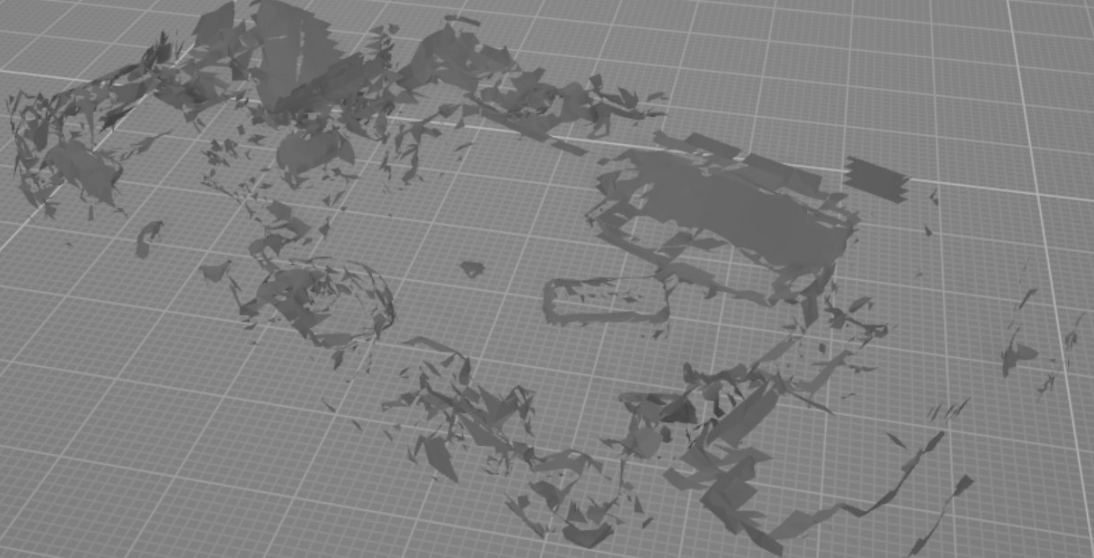}
    \caption{WSA work surface clutter and other higher objects in Home \#1.}
    \label{fig:home1_other_mesh}
\end{figure}

The mesh in the ''other'' category is shown in Figure~\ref{fig:home1_other_mesh}.
This is clutter or objects on the raised work surfaces and the other, higher up objects leftover from the earlier analysis.
These pieces of mesh are included for completeness' sake, and to give an idea to the reader of how the HoloLens emits smaller, higher up surfaces.
Objects such as light fixtures, cabinets, and art pieces are more complex and out of the way.
They are not easily captured by the 3D mesh.

The researcher felt that the objects on raised work surfaces shown in this other mesh were relatively well captured.
Small objects on tables were often overlooked, though.
Whether the HoloLens can accurately capture and be used for evaluating work surface clutter since it is often made of smaller items is an open question.

\subsubsection{WSA OSHA Walking Space Outlines from Home \#1}

\begin{figure}[H]
    \centering
    \includegraphics[width=0.75\columnwidth,keepaspectratio]{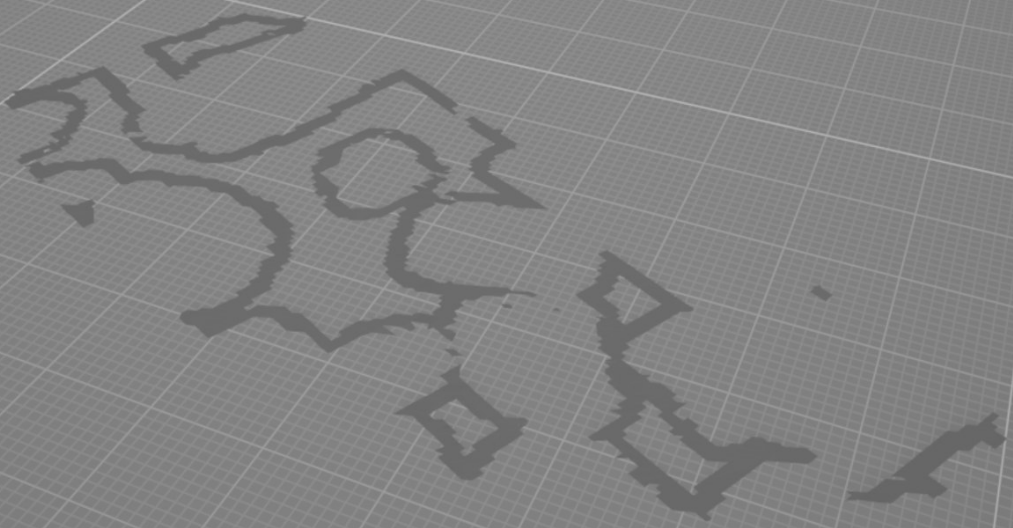}
    \caption{WSA OSHA walking space edges found for Home \#1.}
    \label{fig:home1_osha_edges}
\end{figure}

The final output of the WSA tools is to generate the edges of the OSHA width walking spaces found.
The outlines found for Home \#1 are shown in Figure~\ref{fig:home1_osha_edges}.
These areas the outlines of the locations that had under 0.45m radius of open spaces.
The enclosed sections are then spaces considered ''safe'' walking areas by the WSA algorithm.

When reviewing the ground truth photos and measurement sketches taken during the data collection, the researcher felt that the WSA output was relatively accurate for the homes measured.
The open areas were often narrower than expected, though the open routes seemed to be correct.
More work would be needed to generate quantifed values for the accuracy of the Walking Spaces tool when measuring home living spaces.

\subsubsection{WSA Algorithm Performance}

The lab and apartment runs took around 30 seconds each to process, though the full home samples took several minutes to complete.
This execution was done on a single core 2.1 GHz CPU virtual machine running in a VMWare stack.
This long execution time is likely due to inefficient algorithm implementation in traversing the mesh at each phase due to the larger number of vertices to evaluate.
The waterfall algorithm implementation used here is also very slow for this kind of processing.

\subsection{Visualization Results}

After the WSA tools process the mesh for walking spaces, clutter, and raised surfaces the resulting meshes are used to generate user-facing visualizations.
These renderings use the various WSA classifications to color code a 3D model of the home.
The primary renderings are centered around the availability of walking spaces.
The goal being the presentation of these models to caregivers and clinicians responsible for older adults aging in place.

\begin{figure}[H]
    \centering
    \includegraphics[width=0.75\columnwidth,keepaspectratio]{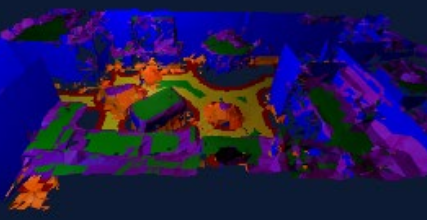}
    \caption{Visualization of walking spaces for Home \#1.}
    \label{fig:home1_visualization}
\end{figure}

In Figure~\ref{fig:home1_visualization}, the various open, OSHA width, walking spaces and the raised work surfaces are colored together.
The yellow and red colors show the floor where it is not a full OSHA width, and a final band around the edge where only about 10cm of floor space is visible.
Orange and purple colors show where clutter is resting on the floor or the ''other'' objects up high reside.

In Home \#1 there are some relatively good walking paths, though there are objects making narrow points in the room.
Notably, the orange colored object is a beanbag chair sitting near a couch with makes the routes in the room too narrow for OSHA standards.
An environmental home assessment would strongly suggest removing the chair in the room to open up safer paths through the space.
The other suggestions would include re-arranging the dining room table behind the couch to open up the walking space from the front door to the middle of the living space.
These are the kinds of objects and furniture arrangements that contribute to the hazards noted in Pynoos, et al.'s summary of environmental hazards~\parencite{Pynoos2010}.

\begin{figure}[H]
    \centering
    \includegraphics[width=0.75\columnwidth,keepaspectratio]{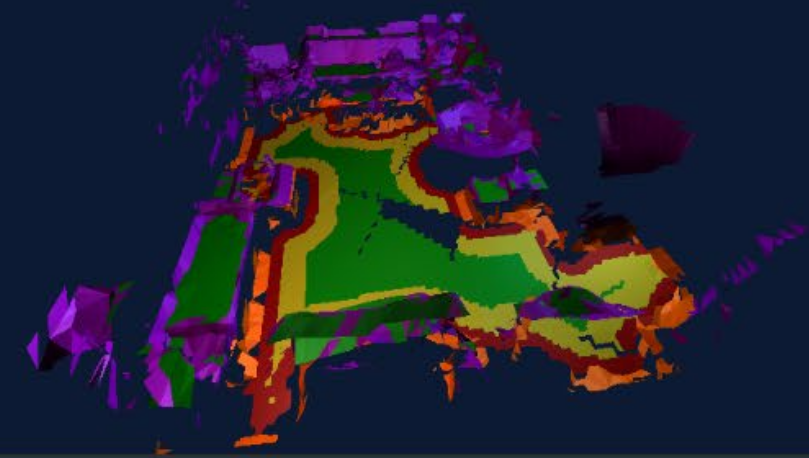}
    \caption{Visualization of walking spaces for apartment \#1.}
    \label{fig:apartment1_visualization}
\end{figure}

In contrast to Home \#1, Apartment \#1's walking spaces were less constricted.
The final visualization, as shown in Figure~\ref{fig:apartment1_visualization}, has a notably larger green area for clear OSHA-width walking spaces.
Residents of the apartment are able to enter through a nearly fully compliant doorway on the right, and reach most of the living space without hazards.

The researcher reviewing the materials and these visualizations was able to quickly compare and contrast the two living spaces for safe walking paths.
Points of restricted or hazardous movement were quickly identified.
The final output aligns with the ground truth images and maps taken from the homes.

Overall, the WSA tools and final visualizations were operational.
The data collection via the HoloLens only took about a minute per room in the homes.
Processing was not real time, but still quick enough for intermittent data collection and rendering for caregivers.
There were notable limitations based around the 3D mesh noise in some lighting and texture situations in the homes, but the overall results were clean and usable as a prototype for future work.

  \section{CONCLUSIONS}
\label{sec:conclusions}

The combination of the 3D mesh generated by the HoloLens v1 and the WSA tools was able to generate accurate annotated visualizations of homes and work spaces.
These tools were able to create images that accurately represented open walking spaces and cluttered areas.
The researcher was able to review the resulting images and map them onto the living spaces they were collected from.

The algorithms worked effectively in homes, apartments, and workspaces.
Objects larger than about 35cm in size were sometimes detected, and all non-black color objects over 50cm were reflected in the mesh.
This should be sufficient for most kinds of clutter detection in day to day living spaces.

This is an early work for these kinds of systems, especially in the in-home healthcare space.
The accuracy of 3D mesh creators like the HoloLens, Structure, or other time of flight sensors is likely to increase in the near future.
With more robust sensing in low light, varying textures, and at higher speeds, these systems should become notably more usable for facilitated environmental home assessments.
The net result should be new tools and approaches for in-home aging in place supporting systems to help older adults live longer and more independently.

\subsection{Future Work}

The Walking Spaces project opens doors to multiple directions for future work.
The first is to continue to refine the mesh interpretation algorithms.
The current approaches to detecting the edges of the clutter and separating out the floors is open to false results with doorways, floor edging, and carpet edges.

The second direction should be in executing a larger, more rigorous data collection for evaluating the tools.
More homes, more accurate measurements of the cluttering objects, and a notation of OSHA and ADA width walking spaces is needed.
With those on hand, the WSA algorithms can be evaluated for accuracy.

Lastly, the resulting visualizations need to be presented to caregivers for user interface design and utility feedback.
Asking potential users what the renderings tell them about the rooms, what more they need to know, and if the information is actionable should be in the immediate next steps for these systems.
  \section{ACKNOWLEDGMENTS}
\label{sec:acknowledgements}

The authors thank Don McMahon from the Washington State University Assistive Technology Research and Development Lab (ATR\&D Lab) for the use of HoloLens v1 equipment and Diane J.~Cook with the Center for Advanced Studies in Adaptive Systems (CASAS) for gerontechnology oriented lab resources.
Special thanks to Team Hekate for their hard work: Konstatin Shvedov, Jacob Stocklass, Jarred Eagley, and Austin Craigie. 

  \section{REFERENCES}
\label{sec:referenced}

{\small
  \printbibliography[heading=none]
}

\end{multicols}

\end{document}